\documentclass[lettersize,journal]{IEEEtran}
\usepackage{amsmath,amsfonts}
\usepackage{cite}
\usepackage{array}
\usepackage[caption=false,font=normalsize,labelfont=sf,textfont=sf]{subfig}
\usepackage{textcomp}
\usepackage{stfloats}
\usepackage{url}
\usepackage{verbatim}
\usepackage{graphicx}
\usepackage[strings]{underscore}
\usepackage[ruled,norelsize,lined,linesnumbered]{algorithm2e}
\usepackage{multirow}
\usepackage{etoolbox}
\usepackage{pifont} 

\makeatletter
\newcommand{\removelatexerror}{\let\@latex@error\@gobble}
\patchcmd{\@maketitle}
  {\addvspace{0.5\baselineskip}\egroup}
  {\addvspace{-1.1\baselineskip}\egroup}
  {}
  {}
\makeatother

\begin{document}
\setlength{\abovedisplayskip}{0pt}
\setlength{\belowdisplayskip}{0pt}
\setlength{\abovedisplayshortskip}{0pt}
\setlength{\belowdisplayshortskip}{0pt}

\setlength{\abovecaptionskip}{0pt}
\setlength{\belowcaptionskip}{0pt}

\setlength{\parskip}{0pt} 

\title{Dynamic Operating Envelopes Embedded Peer-to-Peer-to-Grid Energy Trading}

\author{	Zhisen~Jiang, 
          Ye~Guo,~\IEEEmembership{Senior Member,~IEEE,}
          Hongbin~Sun,~\IEEEmembership{Fellow,~IEEE}
          and~Jianxiao~Wang,~\IEEEmembership{Member,~IEEE}
\thanks{The work was supported in part by the National Natural Science Foundation of China under Grant 51977115. (Corresponding author: Ye Guo, e-mail: guoye@sz.tsinghua.edu.cn)

Zhisen Jiang, Ye Guo, and Hongbin Sun are with the Smart Grid and Renewable Energy Laboratory, Tsinghua-Berkeley Shenzhen Institute, Shenzhen 518071, China.
Hongbin Sun is also affiliated with the State Key Laboratory of Power Systems, Department of Electrical Engineering, Tsinghua University, Beijing 100084, China.

Jianxiao Wang is with the National Engineering Laboratory for Big Data Analysis and Applications, Peking University, Beijing 100871, China.
}}

\maketitle

\begin{abstract}
  A novel decentralized peer-to-peer-to-grid (P2P2G) trading mechanism considering distribution network integrity is proposed. 
  In order to direct prosumers' peer-to-peer (P2P) trading behavior to be grid-friendly, the proposed method incorporates Dynamic Operating Envelopes (DOEs) into the existing P2P2G trading.
  Moreover, DOEs are determined through negotiations between the distribution system operator (DSO) and prosumers alongside the process of P2P trading, avoiding compromising prosumers' privacy and network parameters leakage. 
  To reduce communication costs during P2P trading, a variant of the alternating direction method of multipliers (ADMM), i.e., communication-censored ADMM (COCA) is used to solve the P2P2G trading problem. 
  Finally, the DOE price is shown to be comprised of several economically interpretable components. 
  Simulations validate the effectiveness of the proposed mechanism.
\end{abstract}


\section{Introduction}
\subsection{Background and Motivation}
\IEEEPARstart{I}{N} recent years, the penetration of distributed energy resources (DERs) has grown rapidly, especially for renewable energy sources (RESs). 
Thus, traditional passive consumers with these active assets are becoming prosumers. 
Under the most dominant renewable energy compensation scheme based on Feed-in-Tariff (FiT) and Time-of-Use (ToU), these prosumers can make decisions to sell or buy power from the distribution network at given prices, which is called P2G energy trading. 
As a supplement to the existing P2G scheme, P2P trading emerges, allowing prosumers to transact energy with each other in a platform via bilateral contracts anonymously and autonomously. 
Typically, prosumers may set P2P prices between FiT and ToU to make P2P trading more beneficial than their transactions with the grid.

\subsection{Related Works}
Different decentralized methods for solving P2P trading problems have been proposed.
In these methods, P2P problems are usually modeled as non-cooperative games \cite{GNE}, 
or formulated as a social welfare maximization problem and solved by distributed algorithms,
e.g. consensus-based ADMM \cite{ADMM1, ADMM2, DC_ADMM, Online_ADMM} or Relaxed Consensus+Innovation \cite{RCI}.
Although above decentralized methods preserve the privacy of prosumers, they cannot consider the network integrity in detail.
Network integrity in this paper refers to nodal voltage magnitudes and line powers are within safe ranges in the distribution network.
Authors of \cite{impact_overview, impact_overview2,Tushar_overview} demonstrate the impact of P2P energy transactions on the distribution network, concluding that P2P trading may cause voltage limit violations at prosumers' nodes and increase the overall network loss.

To ensure network integrity during P2P trading, centralized approaches have also been proposed.
These approaches require the DSO to directly intervene in participants' trades, including blocking trades and assigning P2P matching to prevent potential network constraint violations.
Reference \cite{Decentral_Guerrero} enables the DSO to do network sensitivity analysis after gathering each pair of prosumers' P2P trading contracts, after which it rejects ones that may jeopardize network integrity.
Authors of \cite{distance_match_Guerrero} present an electrical-distance-based P2P matching mechanism, which matches those prosumers located physically close in priority to reduce line power overloading. 
However, these approaches rely on the DSO to directly manage P2P trades, thus contradicting the requirement for privacy preservation of trade information.

To integrate the strengths of the two previously mentioned techniques, designing integrity-informed decentralized methods for P2P trading decisions is an effective approach.
We can apply distributed algorithms to incorporate network constraints into prosumers' decision models.
Reference \cite{Distribute_constraints_Ullah+GUP} uses a fast ADMM (F-ADMM) method to add network constraints into prosumers' trading models.
Authors of \cite{Distribute_constraints_Zhong, Distribute_constraints_Jia} formulate the P2P trading as a Nash Bargaining model considering network constraints to ensure voltages and currents are within permitted ranges. Then several distributed algorithms are used to solve it.
However, these papers simply use the linearized power flow model to characterize the distribution network, neglecting the increased power loss caused by unexpected P2P trades.
To improve the accuracy, reference \cite{Distribute_constraints} uses sensitivity analysis to evaluate the change of nodal voltage and network power loss arising from P2P trading, followed by a fast dual ascent method to internalize operational constraints into prosumers' trading decision problems.
The drawbacks of these methods are prosumers are actually not willing or capable to consider network constraints during the trading stage, which has nothing to do with their individual surplus.
Moreover, it is also impractical for the DSO to share network parameters with prosumers.

Apart from applying distributed algorithms to incorporate network constraints, during the trading process, the DSO can ask prosumers to pay the corresponding grid utilization charge (GUC) for their P2P trades, thus directing trades to be grid-friendly.
In \cite{Distribute_constraints_Ullah+GUP, Distance_Paudel2, Distance_Paudel1, Exo_Baroche}, an electrical distance-based GUC is presented to charge each pair of prosumers.
Reference \cite{DLMP_Dominant_GUP, DLMP_Zhang, DLMP_multi_round} intend to associate P2P trading interactions with distribution network operations, computing GUC through the difference of distribution locational marginal price (DLMP) between two prosumers' nodes.
One of the major drawbacks of these DLMP-based GUC mechanisms is that these prosumer-specific DLMPs are broadcasted to all entities at each round of P2P trading negotiation, which compromises the anonymity of prosumers' participation in P2P trading.
In addition, as GUC is announced before trades happen, these methods cannot consider the DLMP's variation with amounts of P2P trades.

While the mentioned P2P trading models effectively address network constraints, they typically optimize for a single power operating point among P2P participants to ensure network integrity.
The more general and easy-to-implement approach is to optimize the pre-defined time-variant network access limits for P2P participants, which has been realized in practical use as Dynamic Operating Envelopes.
DOE is defined as “operating envelope that varies export limits over time and location based on the available capacity of the local network or power system as a whole” in \cite{DEIP_outcome}. 
Once DOEs are calculated and allocated to different prosumers, voltages will not exceed permissible bounds when net injections of all prosumers are within the DOEs.
However, traditional DOEs calculation methods in \cite{DOE_ensuring, DOE_uncertainty, DOE_tap, DOE_fair} do not consider P2P tradings, which means they only limit prosumers' P2G behaviors.
Extending the calculation of DOE to further apply it to P2P trading is promptly needed.

In the course of composing this paper, the investigations presented in \cite{DOE_P2P1} and \cite{DOE_P2P2} have undertaken preliminary efforts to integrate DOEs into P2P trading markets. 
These studies employed a direct incorporation of DOE calculations prior to P2P trading negotiations, utilizing the centralized DOE calculation methodology as proposed in \cite{DOE_ensuring}. 
However, these approaches overlook the incorporation of trading intentions into DOE calculations, thus neglecting the potential for adapting trading strategies to alter prosumers' export limit requests. 
Furthermore, the method outlined in \cite{DOE_ensuring} assumes comprehensive observability of prosumers' DER information by the DSO for improved network capacity utilization in DOE calculations, thereby compromising prosumers' privacy. 
In light of the inherent flexibility of P2P trading mechanisms, there is an imperative need for the development of a novel DOE calculation framework that accounts for P2P trading, alongside an exploration of decentralized DOE computation methodologies.

Furthermore, existing literature on DOE calculation predominantly features the direct issuance of export limits by the DSO according to prosumers' asks. 
An inherent concern arises, wherein prosumers may provide untruthful requests for their intended export limits. 
In response to this challenge, some work has undertaken preliminary inquiries into the pricing mechanisms associated with DOE allocation. 
Notably, reference \cite{ChenCong} introduces the concept of \textit{Locational Allocation Price}, yet lacks specificity regarding the DSO's cost function and overlooks the P2P2G trading. 
This highlights the imperative for comprehensive explorations of such pricing schemes.
\vspace{-1.0em}
\begin{table}[h]
\vspace{-1.0em}
\caption{Comparison between this paper and others.} \label{tab:review}
\resizebox{\columnwidth}{!}{%
\begin{tabular}{ccccc}
\hline
Ref. & Network Constraints & P2P Trading & \multicolumn{1}{l}{DOE Calculation} & \multicolumn{1}{l}{DOE Price} \\ \hline
\cite{Decentral_Guerrero,distance_match_Guerrero}          & Centralized approach   & \ding{52} & \ding{56}           & \ding{56} \\
\cite{Distribute_constraints_Ullah+GUP,Distribute_constraints_Zhong, Distribute_constraints_Jia,Distribute_constraints}         & Distributed algorithms & \ding{52} & \ding{56}           & \ding{56} \\
\cite{Distribute_constraints_Ullah+GUP, Distance_Paudel2, Distance_Paudel1, Exo_Baroche,DLMP_Dominant_GUP, DLMP_Zhang, DLMP_multi_round} & GUC                    & \ding{52} & \ding{56}           & \ding{56} \\
\cite{DOE_ensuring, DOE_uncertainty, DOE_tap, DOE_fair}          & DOE                    & \ding{56} & Centralized   & \ding{56} \\
\cite{DOE_P2P1, DOE_P2P2}             & DOE                    & \ding{52} & Centralized   & \ding{56}   \\ 
\cite{ChenCong}             & DOE                    & \ding{56} & Centralized   & \ding{52}   \\ \hline
This paper           & DOE                    & \ding{52} & Decentralized & \ding{52}   \\ \hline
\end{tabular}%
}
\end{table}
Table \ref{tab:review} provides a summary of the features of the relevant existing
papers reviewed above, highlighting the differences between this paper and others regarding to the way to consider network constraints, P2P trading, DOE calculation method and DOE prices.
\vspace{-1.0em}
\subsection{Contributions}
In this paper, the DOE is embedded in the P2P2G trading mechanism to ensure network integrity.
To take injection caused by P2P trades into consideration, this paper modifies the calculation of DOEs.
Also, the DOE is determined through the negotiation between the DSO and prosumers.
Therefore, prosumers' privacy is preserved and network integrity is guaranteed.

Main contributions are summarized as follows.

1) By incorporating DOE into the P2P2G trading, a novel decentralized trading mechanism is proposed to keep distribution network integrity.
Because P2P trading amounts are contained in the net power injection of prosumer-specific nodes, prosumers' trading behaviors can be directed to be grid-friendly through the DOE issued by the DSO, which is easy to implement in the existing distribution network.

2) In the proposed mechanism, DOEs are determined through the negotiation between the DSO and prosumers based on ADMM, simultaneously alongside the P2P trading process among prosumers.
The DOE negotiation is bilaterally privacy-preserving by establishing a clear separation between prosumers' privacy information (i.e. P2P trading amounts and devices parameters) and any grid parameters owned by the DSO.
For maximizing individual surplus, prosumers will adjust their P2P trades to satisfy DOEs or pay for enlarging DOEs to allow more P2P trading amounts in each round of negotiation.

3) A COCA technique is leveraged to solve the trading problem, reducing the cost of communication for prosumers to the largest extent.

4) An interpretable DOE price is derived by duality analysis, which cannot only be utilized as a kind of price signal through DOE negotiation but also as a component of P2P trading price.

The remainder of this paper is organized as follows: 
Section \ref{sec:framework} introduces DOE, the entities envolved and the P2P2G framework. 
Section \ref{sec:models} formulates the DOE calculation model with P2P2G trading, prosumers models and the DSO's model.
Section \ref{sec:reformulation_sol} gives the solution algorithm.
Also, we elaborate the DOE negotiation and P2P2G trading process in it.
Section \ref{sec:case} shows case studies and section \ref{sec:conclusion} concludes this paper.
\section{Preliminary and Framework}\label{sec:framework}
In this section,  we first provide a conceptual overview of DOE.
Then we introduce roles of different entities in the proposed mechanism and describe the overall framework.
\subsection{Concept of Dynamic Operating Envelope}
\begin{figure}[htb]
  \centering
  \includegraphics[width=0.95\columnwidth,scale=0.5]{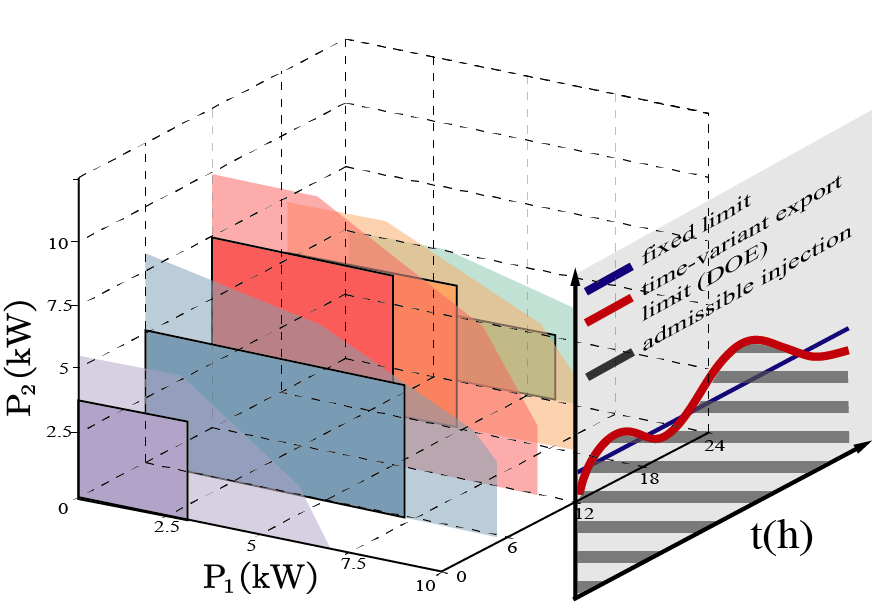}
  \caption{The Dynamic Operating Envelope.}
  \label{fig:DOE}
  \end{figure}

DOE provides a specific feasible range for each prosumer's power injections while considering the DSO's objective for preserving the network integrity by ensuring voltages and line power are within permissible bounds.
As illustrated in Fig. \ref{fig:DOE}, the feasible region (FR), represented as a polytope,varies during different time intervals due to diverse network operational states. 
These variations may arise from factors such as time-variant demands at nodes where no prosumer is located.
As it is hard to explore the exact high-dimensional FR, the DSO can internally allocate it to different prosumers (rectangles) according to a given objective function.
The combination of different time allocations forms the admissible power injection range for a prosumer (shadow).
In doing so, prosumers can operate independently when their power injections are within their given DOEs.

In this paper, we adopt the same approach as described in \cite{DOE_uncertainty, AdvancedVPP} where only use one-side DOE to limit prosumers' export behaviors, because the reverse power flow arising from too much export power usually causes more problems in a distribution network. 
The envelope for the import limit can be solved using the same method in this paper.
Based on this assumption, the key to obtaining DOEs is to solve the time-variant export limits $\boldsymbol{P}_{i}^{e} = \left\{ P_{i,t}^{e} \right\} _{t\in \mathcal{T}}$ (red curve) for the $i$-th prosumer, given selected time horizon $\mathcal{T} \,\,=\,\,\left\{ 1,\dots ,T \right\}$.
Then, the injection region given as $\left\{ \left[ -P_{i,t}^{imp},P_{i,t}^{e} \right] \right\} _{t\in \mathcal{T}}$ for all $i \in \mathcal{N}$ is formed and sent to the $i$-th prosumer, where $\mathcal{N} \,\,=\,\,\left\{ 1,\dots ,N \right\}$ is the set of indices for prosumers. 
$P_{i,t}^{imp}$ represents a predetermined power import limit. Since there are no constraints on power import behavior, this limit can be significantly greater than $P_{i,t}^{e}$ and is not treated as a variable subject to optimizations in this paper.
\subsection{Entities Description}
There are two kinds of entities involved in the problem of P2P2G trading, i.e., prosumers and the DSO.
Prosumers located at distribution networks may be equipped with RESs, energy storage systems and non-flexible loads. 
They can sell or buy energy from the distribution network at FiT and ToU respectively, as well as trade energy with each other in a P2P platform.
Their objectives are maximizing individual surplus under operational constraints. 

Another important role in P2P2G trading is the DSO. 
The DSO is obligated for DOE calculation whose objective is to minimize the operation cost of the distribution network under the power injection pattern associated with a DOE allocation.
We also assume the DSO does not know each P2P trading amount and the price.
Through the introduction of DOEs, there is no need for the DSO to decide or monitor each P2P trade, which separates the DSO from P2P trading. 
\subsection{P2P2G Framework}
\begin{figure*}[htb]
  \centering
  \subfloat[]{\includegraphics[width=0.95\columnwidth]{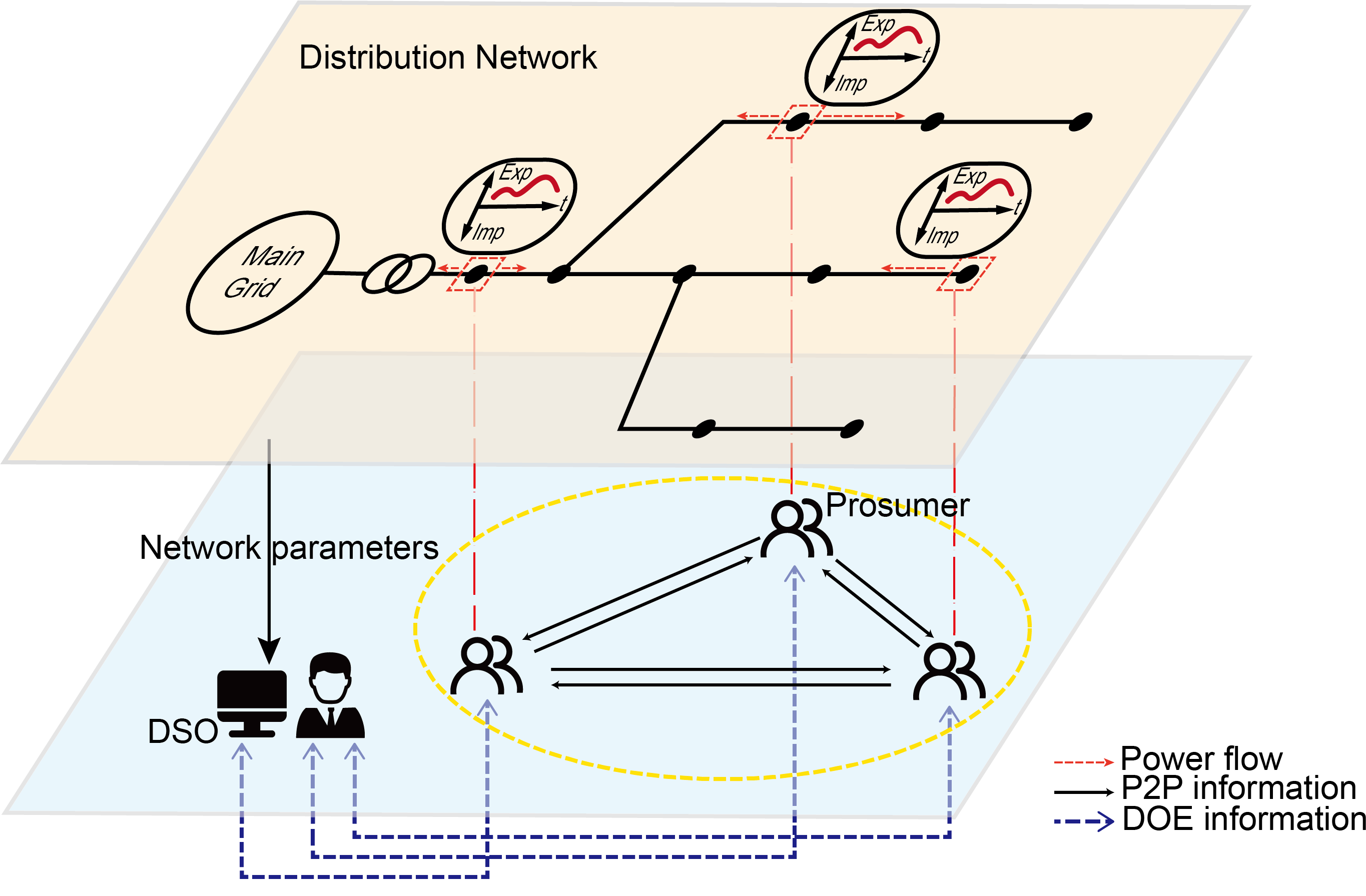}%
  \label{fig:framework}}
  \hfil
  \subfloat[]{\includegraphics[width=0.95\columnwidth]{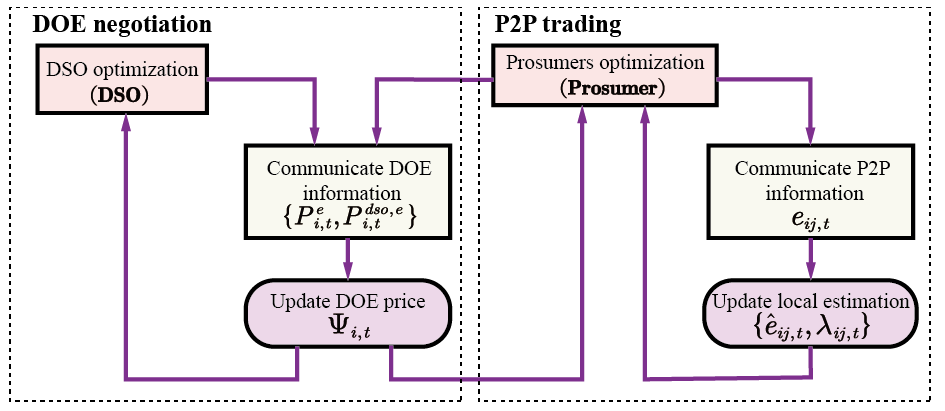}%
  \label{fig:loops}}
  \caption{Overview of P2P2G market mechanism and negotiation process: (a) Framework. (b) Negotiations.}
  \label{fig1}
  \end{figure*}

In this work, we focus on the day-ahead market, without considering the uncertainty of RES outputs and loads. 
It is also assumed that each prosumer is aware of others' trading requests.
The overview of the proposed mechanism is given in Fig. \ref{fig1}(a). 
Therein, prosumers need to determine their desired DOE and P2P trades together, then negotiate with the DSO and other peers.
When both prosumers and DSO reach a consensus, negotiations terminate. 
As shown in Fig. \ref{fig1}(b), the negotiation process includes two loops:

1) \textit{Left loop:} This loop describes the DOE negotiation process.
While trading with peers, prosumers need to submit their intended export limits to the DSO. 
The first round submission of the initial intended export limits is based on a preset DOE price from the DSO.
When the DSO gathers all prosumer-side export limits, it will run an Optimal Power Flow (OPF) to decide DSO-side export limits.
Thanks to the use of ADMM, the multiplier for DOE reciprocity constraints can be updated and broadcast to prosumers, regarded as DOE prices.
Corresponding to the DSO's feedback, prosumers will change their P2P2G trading amounts to meet the last-round DOE allocation.

2) \textit{Right loop:} Another loop is mainly about the P2P trading process.
In this loop, prosumers determine their intended P2G trading amounts, P2P trading amounts and export limit, accounting for the last round DOE allocation and DOE price issued by the DSO.
Then, each prosumer exchanges its P2P trading amounts with its partners.
Note that COCA is used to solve the optimization, prosumers will not need to send the updated trading amounts if they are not sufficiently different from the previous one.
After receiving all peers' trading amounts, the prosumer will update trading amounts and trading prices locally.

It is crucial to emphasize that these two loops influence each other.
For prosumers, they intend to adjust their operations and P2P2G trading amounts in each negotiation following the newest DOEs issued by the DSO.
Thus, the DSO guarantees network integrity via DOEs, without directly monitoring P2P trading amounts.   

\section{Prosumers, DSO and DOE calculation Model with P2P2G Energy Trading}\label{sec:models}
In this section, we first introduce a simplified symbolic representation of the DOE calculation model with P2P2G trading. 
Then we illustrate the prosumers' and the DSO's models.
Elaborate model decomposition and the solution algorithm will be expounded upon in the ensuing sections.
\subsection{DOE calculation model with P2P2G trading}
The determination of DOE is based on the methodology presented in \cite{ChenCong}.
Better than \cite{ChenCong}, the model is incorporated with the P2P2G trading process.
The DOE calculation model with P2P2G trading can be formulated as follows:
\begin{flalign}
  &(\mathrm{I})\underset{\left\{ P_{i,t}^{e},\forall i\in \mathcal{N} ,t\in \mathcal{T} \right\}}{\mathrm{minimize}}\,\,J\left( \left\{ \boldsymbol{P}_{i}^{e} \right\} _{i=1}^{N} \right) +\sum_{i\in \mathcal{N}}{\varphi _i\left( \boldsymbol{P}_{i}^{e} \right)}\notag&
  \\
  &\mathrm{subject}\ \mathrm{to}& \notag
  \\
  &\mathrm{For} \ \mathrm{all}\ i\in \mathcal{N}:\varphi _i\left( \boldsymbol{P}_{i}^{e} \right) =\underset{x_i\in \mathcal{X} _i\left( \boldsymbol{P}_{i}^{e} \right)}{\mathrm{minimize}}\,\,\theta _i\left( x_i \right) \label{doe_prosumers}&
  \\
  &\forall \boldsymbol{p}_{i}^{inj}\in \left[ -\boldsymbol{P}_{i}^{imp},\boldsymbol{P}_{i}^{e} \right] :x_{net}\in \mathcal{X} _{net}\left( \left\{ \boldsymbol{p}_{i}^{inj} \right\} _{i=1}^{N} \right)  \label{doe_inj} &
\end{flalign}
where $J\left( \cdot \right): \mathbb{R} ^{NT}\rightarrow \mathbb{R} ,\varphi _i\left( \cdot \right) : \mathbb{R} ^{T}\rightarrow \mathbb{R} $ represents the cost of the DSO for providing export limits and the $i$-th prosumer's cost for P2P2G trading if it is given such export limits.
$x_i$ and $\theta_i \left( \cdot \right)$ are the variables and cost function for the $i$-th prosumer.
The feasible region of the prosumer's P2P2G trading problem, denoted as $\mathcal{X}_i\left( \boldsymbol{P}_{i}^{e} \right)$, is determined based on the allocated export limits, as implied by constraint (\ref{doe_prosumers}).
$\mathcal{X} _{net}\left( \left\{ \boldsymbol{p}_{i}^{inj} \right\} _{i=1}^{N} \right) $ is the feasible region of the distribution network, which is represented by the power flow equations, bounded voltage and line power constraints.
Constraint (\ref{doe_inj}) enforces the robustness of the model, ensuring that if any power injection of prosumers falls within the provided $\left[ -\boldsymbol{P}_{i}^{imp},\boldsymbol{P}_{i}^{e} \right]$, the integrity of the distribution network is maintained. 

The solution to (I) yields the time-variant export limits for prosumers (DOE), denoted as $\left\{ \boldsymbol{P}_{i}^{e,*} \right\} _{i=1}^{N}$, and $\left\{ \left[ -{P}_{i,t}^{imp},{P}_{i,t}^{e,*} \right]  \right\} _{t=1}^{T} $ is issued to the $i$-th prosumer.
The robust constraint (\ref{doe_inj}) can be further simplified into:
\begin{equation}
x_{net}\in \mathcal{X} _{net}\left( \left\{ \boldsymbol{P}_{i}^{e} \right\} _{i=1}^{N} \right) \label{X_net}
\end{equation}
This transformation can be justified for a balanced radial network, where more injections from nodes will result in higher currents running towards the root of the network, which in turn leads to higher voltages along the network lines.
So when $\boldsymbol{P}_{i}^{e,*}$ is permissible for the network, any injection within $\left[ -\boldsymbol{P}_{i}^{imp},\boldsymbol{P}_{i}^{e,*} \right]$ is also permissible \cite{Liu_sensitivity}.

Notice that constraint (\ref{doe_prosumers}) incorporates the P2P2G trading problem of prosumers.
Subsequent sections will elucidate the P2P2G trading problem among prosumers, incorporating DOE negotiations between prosumers and the DSO to solve (I).
\vspace{-1.0em}
\subsection{Prosumers models}
\begin{figure}[t]
  \centering
  \includegraphics[width=0.95\columnwidth]{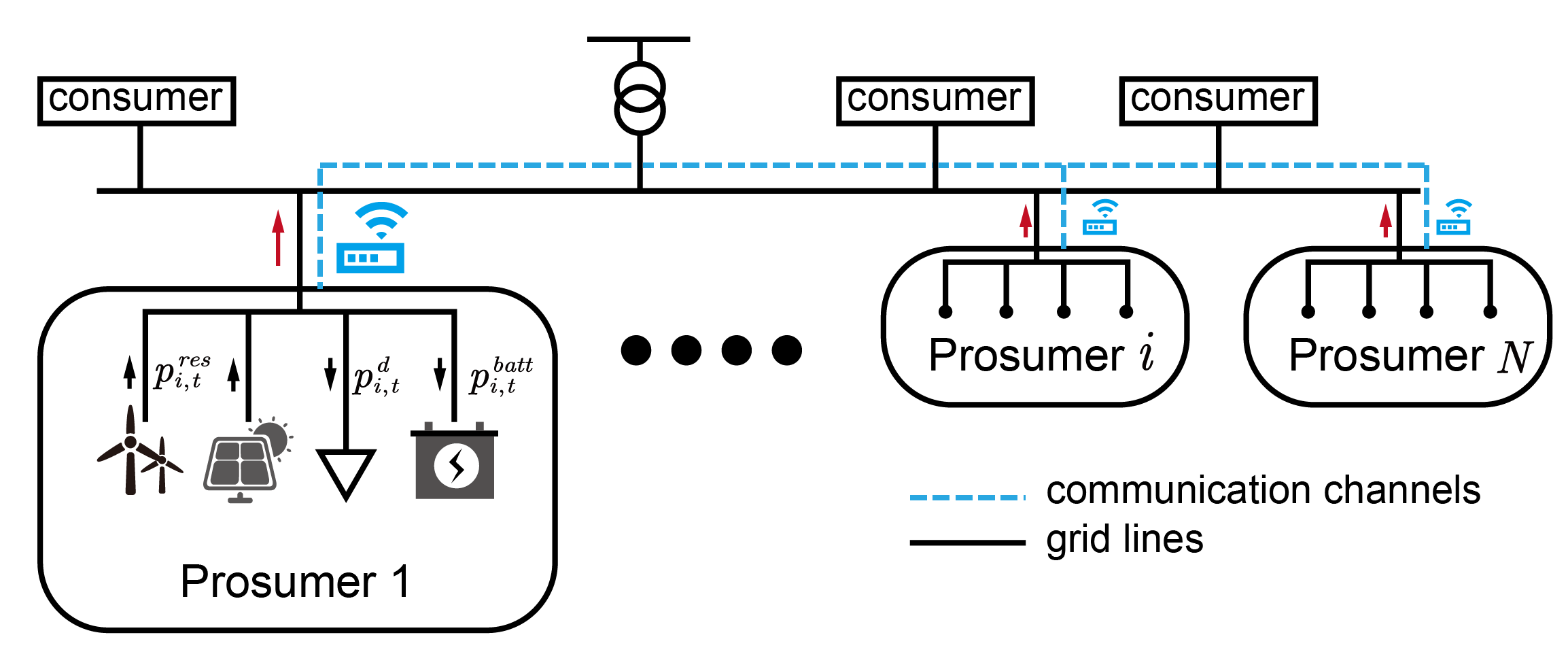}
  \caption{Prosumers' model.}
  \label{fig:prosumers}
  \end{figure}

Without loss of generality, we acknowledge the following assumptions in our study: 
\begin{enumerate}
  \item As shown in Fig. \ref{fig:prosumers}, each prosumer is equipped with RESs and energy storage, and they connect with each other through the distribution network, with each node accommodating only one prosumer.
  \item Each prosumer is capable to derive the optimal operation and trading plan by solving its local problem.
  \item For simplicity of the following price properties statement, we ignore the charging and discharging efficiency of storage.
  \item Consistent with policies, prosumers are prohibited from buying or selling power at the same time with the distribution network.
\end{enumerate} 

The unordered pair $(i,j) \in \mathcal{E}$ means agents $i$ and $j$ can trade with each other, where $\mathcal{E}$ is the set of possible trading pairs. The set of trading partners of prosumer $i$ is defined as $\mathcal{N} _i\,\,=\,\,\left\{ j|\left( j,i \right) \in \mathcal{E} \right\} $.
Next, we present the objective function and detailed constraints for each prosumer's decision-making model. Greek alphabets given in brackets before each constraint are Lagrange multipliers.

1) \textit{Objective function}: A prosumer's objective is to maximize the total surplus in such P2P2G trading. 
So the objective function is given by (for minimization):
\begin{equation}
  \label{obj_agent}
  \theta _i=\sum_{t\in \mathcal{T}}\Delta T{\left( \lambda _{t}^{+}p_{i,t}^{+}-\lambda _{t}^{-}p_{i,t}^{-}-\sum_{j\in \mathcal{N} _i}{\lambda_{ij,t}e_{ij,t}} \right)}
\end{equation}
where $\lambda _{t}^{+}$, $\lambda _{t}^{-}$ and $\lambda_{ij,t}$ mean ToU, FiT and P2P price respectively.
$p_{i,t}^{+}$ and $p_{i,t}^{-}$ denote the power of purchase and sale with the distribution network, and $e_{ij,t}$ is the power selling from prosumer $i$ to $j$.
The first two terms represent the cost of P2G trading. 
The last term gives the P2P trading revenue, $\lambda_{ij,t}$ is agreed by trading partners through the negotiation process, which will be discussed later in detail.

2) \textit{Battery constraints:} Each prosumer may own local energy storage.
The storage system is restricted by the following constraints:
\begin{align} 
  &(\sigma^{-} _{i,t},\sigma^{+} _{i,t}): \underline{p}_{i}^{batt}\le p_{i,t}^{batt}\le \overline{p}_{i}^{batt} \label{batt}\\
  &(\rho^{-} _{i,t},\rho^{+} _{i,t}): \underline{E}^{s}_{i} \le E^{s}_{i,t}\le \overline{E}_{i}^{s} \label{E}\\
  &E^{s}_{i,t+1}=E^{s}_{i,t}+p_{i,t}^{batt}\Delta T \label{Eb}
\end{align}
where $p_{i,t}^{batt}$ and $E^{s}_{i,t}$ denote a prosumer's battery net charging power and measured state of charge (SoC) respectively.
For simplicity, we do not split the battery power into charging and discharging.
$\underline{p}_{i}^{batt}, \overline{p}_{i}^{batt}, \underline{E}_{i}^{s}, \overline{E}_{i}^{s}$ are rated charging power and measured SoC limits of the storage. 
Constraint (\ref{Eb}) gives the relationship between measured SoC and net charging power. 
This model is simplified and does not account for the different modes of charging and discharging.

3) \textit{Local power balance constraint:} The local power balance of each prosumer $i \in \mathcal{N}$ is represented by the following equation:
\begin{align}
  &\phi_{i,t}: p_{i,t}^{+}-p_{i,t}^{-}=p_{i,t}^{d}-p_{i,t}^{res}+p_{i,t}^{batt}+p_{i,t}^{P2P} \label{pb}\\  
  &\mu_{i,t}: p_{i,t}^{P2P}\,\,=\sum_{j\in \mathcal{N} _i}{e_{ij,t}} \label{ps}
\end{align}
where $p_{i,t}^{d}$ and $p_{i,t}^{res}$ denote the prediction of non-flexible local power demand and RES output respectively.
$p_{i,t}^{P2P}$ denotes the total P2P trading amounts of prosumer $i$.
In this work, we avoid the tedious work of assigning buyers and sellers beforehand\cite{Online_ADMM,Asychron_ADMM,RCI}. 
Instead, the role of prosumers in the P2P market is determined by the sign of $p^{P2P}_{i,t}$. 
If $p^{P2P}_{i,t} \ge 0$, prosumer $i$ is viewed as a seller at time $t$, but it can alter into a buyer in another time interval.

4) \textit{P2G trading constraints:} Prosumers' P2G behaviors are limited by:
\begin{align}
  & 0\le p_{i,t}^{+}\le \overline{p^{+}}z_{i,t}\\
	& 0\le p_{i,t}^{-}\le \overline{p^{-}}(1-z_{i,t})
\end{align}
where $z_{i,t}$ are binary variables characterizing P2G trading mode. 
$\overline{p^{+}}, \overline{p^{-}}$ are trading limits.
However, these constraints can be exactly relaxed into:
\begin{align}
  &(\alpha^{-} _{i,t},\alpha^{+} _{i,t}): 0\le p_{i,t}^{+}\le \overline{p^{+}} \label{purchase}\\
	&(\beta^{-} _{i,t},\beta^{+} _{i,t}):  0\le p_{i,t}^{-}\le \overline{p^{-}} \label{sell}
\end{align}
We provide a concise overview of the exact relaxation method. 
Initially, we consider $p_{i,t}^{+}$ and $p_{i,t}^{-}$ to be strictly greater than zero. 
By applying the Karush-Kuhn-Tucker (KKT) condition to the prosumers' models discussed in Section \ref{sec:reformulation_sol}, we observe a contradiction with the aforementioned assumption. 
Therefore, we conclude that the relaxation is exact, implying that $p_{i,t}^{+}$ and $p_{i,t}^{-}$ cannot both be larger than zero simultaneously.

5) \textit{P2P trade constraint:} The trading constraint is given by:
\begin{equation}
    \lambda _{ij,t}:e_{ij,t}=-e_{ji,t} \label{recipro0}
\end{equation}
For each P2P trade, it must satisfy the constraint (\ref{recipro0}), where $e_{ij,t}$ is the decision variable of prosumer $i$, while $e_{ji,t}$ is that of partner $j$. 
This constraint is called reciprocity constraint, indicating the trading agreement of each pair of prosumers.
Moreover, the multiplier of the reciprocity constraint is taken as the P2P trading price.

6) \textit{DOE constraint:} Focusing on the connection point where a prosumer trades with other peers, the net power injection can be split into two parts: 
interaction with the distribution network (P2G behavior) and power traded with peers (P2P behavior).
\begin{equation}
  \epsilon _{i,t}: p_{i}^{inj}=p_{i}^{-}-p_{i}^{+}+p_{i}^{P2P} \label{inj}
\end{equation}
According to the definition of DOE, each prosumer should restrict their net power injection below the given export limit (with the assumption that the import limit $P_{i,t}^{imp}$ is not exceeded):
\begin{gather}
  \gamma_{i,t}:p_{i}^{-}-p_{i}^{+}+p_{i}^{P2P}\le P_{i,t}^{e} \label{prosumer_DOE}
\end{gather} 

\textit{Remark 1:} Traditional DOE representation excludes P2P trading power injection, thus ignoring the P2P trading effects on the network \cite{DOE_ensuring, DOE_uncertainty, DOE_tap, DOE_fair}.
By (\ref{prosumer_DOE}), network integrity is taken into consideration when prosumers trade with each other.
$P_{i,t}^{e}$ is taken as a variable, which means prosumers can adjust their P2P trades to change their intended DOE ask.
\vspace{-1.0em}
\subsection{DSO's model}
\vspace{-0.5em}
In the following, we first use implicit functions to model network constraints, in which network state variables like voltage magnitudes are not directly expressed as decision variables.
Instead, they are functions of each prosumer node's export limits and fixed loads at other nodes.
We prefer to use the implicit function because, in later analysis, the so-called DOE price can admit an economical interpretation.

Consider a tree network $\mathcal{G} ^n\,\,=\,\,\left( \overline{\mathcal{N}} ,\mathcal{L} \right) $, where $\overline{\mathcal{N}}=\left\{ 0,1,\dots ,\overline{N} \right\}$is the set of all nodes, $\mathcal{L}=\left\{ 1,\dots ,\overline{N} \right\}$ gives the set of lines.
Setting root node connected with the upstream grid (node 0) as the reference, $\overline{\mathcal{N}}^{+}=\left\{1,\dots ,\overline{N} \right\}=\mathcal{N}\cup \mathcal{K}$, where $\mathcal{K}$ is the set of nodes where no prosumer is located.
Take the prediction of reactive power injection $q_{i,t}^{d}$ for $i\in \overline{\mathcal{N}}^{+}$, prediction of active power injection $p_{i,t}^{d}$ for $i\in \mathcal{K}$ and voltage magnitude of the root node $v_0$ as the input, the distribution network constraints are:
\begin{flalign}
  &\mathrm{For}\ \mathrm{all}\ t\in \mathcal{T} ,j\in \mathcal{L} , i\in \mathcal{\overline{N}^{+}}:\notag &\\
  &\eta _{j,t}:f_{j,t}^{p}(P_{i,t}^{e},v_0,\boldsymbol{y})^2+f_{j,t}^{q}(P_{i,t}^{e},v_0,\boldsymbol{y})^2\le S_{j}^{2} \label{s1}&\\
  &\Delta _{j,t}:[f_{j,t}^{p}(P_{i,t}^{e},v_0,\boldsymbol{y})-R_jl_{j,t}(P_{i,t}^{e},v_0,\boldsymbol{y})]^2& \notag\\
  &\quad \quad +[f_{j,t}^{q}(P_{i,t}^{e},v_0,\boldsymbol{y})-X_jl_{j,t}(P_{i,t}^{e},v_0,\boldsymbol{y})]^2\le S_{j}^{2} \label{s2}&\\
  &(\tau^{-}_{i,t},\tau^{+}_{i,t}):\underline{v}_{i}\le v_{i,t}(P_{i,t}^{e},v_0,\boldsymbol{y})\le \bar{v}_i \label{v}&\\
  &\Omega _{p,t}:\sum_{i\in \mathcal{N}}{P_{i,t}^{e}}+\sum_{k\in \mathcal{K}}{p_{k,t}^{d}}+\sum_{j\in \mathcal{L}}{R_jl_{j,t}(P_{i,t}^{e},v_0,\boldsymbol{y})=p_{0,t}} \label{p}&\\
  &\Omega _{q,t}: \sum_{i\in \mathcal{\overline{N}^{+}}}{q_{i,t}^{d}}+\sum_{j\in \mathcal{L}}{X_jl_{j,t}(P_{i,t}^{e},v_0,\boldsymbol{y})=q_{0,t}} \label{q}&
\end{flalign}
where $\boldsymbol{y} = \{\{q_{i,t}^{d}\}_{i\in \mathcal{\overline{N}^{+}}}, \{p_{i,t}^{d}\}_{i\in \mathcal{K}}\}$.
$f_{j,t}^{p,q},l_{j,t}$ are the line active power, reactive power, and square of the current magnitude at $t$.
$v_i$ is the voltage magnitude.
$p_{0,t},q_{0,t}$ are active and reactive power injections at the root node at $t$.
Constraints (\ref{s1}) and (\ref{s2}) limit lines power within range, and constraint (\ref{v}) limits voltage magnitudes.
Power balance is ensured by constraints (\ref{p}) and (\ref{q}).
Constraint (\ref{s1})-(\ref{q}) depicts the $\mathcal{X} _{net}\left( \left\{ \boldsymbol{P}_{i}^{e} \right\} _{i=1}^{N} \right)$ in (\ref{X_net}), which indicates that when power injections are the same as export limits, the voltage and line power are in the admissible range.
Further, the power injections in the envelope $\left\{ \left[ -P_{i,t}^{imp},P_{i,t}^{e} \right] \right\} _{t\in \mathcal{T}}$ for all $i \in \mathcal{N}$ also ensure the network integrity.

The cost incurred by the DSO to provide export limits, denoted as $J\left( \left\{ \boldsymbol{P}_{i}^{e} \right\} _{i=1}^{N} \right)$, is defined as $\frac{1}{S}\sum_{s\in S}{\sum_{t\in \mathcal{T}}{\sum_{j\in \mathcal{L}}{\pi _tR_jl_{j,t,s}\left( \left\{ \frac{s}{S}P_{i,t}^{e} \right\} _{i=1}^{N} \right)}}}\Delta T$ in this paper, where $\pi_{t}$ is the energy price at the root node of the distribution network.
We build $J$ by following the approach introduced in \cite{DOE_price}, wherein the authors partition the envelope interval into uniform segments and determine the loss cost when all prosumers' power injections occur simultaneously within the same segment.
Here, $\mathcal{S}=\{1,2,\cdots,S\}$ is the set of indices for scenarios, with $S$ being the total number of given scenarios. 
In this context, scenario $s$ indicates a power injection at $\frac{s}{S}$ of the export limit, and $l_{j,t,s}$ denotes the corresponding square of the current magnitude.
The defined cost function represents the expectation of the network loss cost when power injections are below the specified export limits.
\vspace{-1.0em}
\section{Model Decomposition and Solution Algorithm}\label{sec:reformulation_sol}
\vspace{-1.0em}
In this section, we first reformulate (I) and use ADMM to decompose it into DSO's DOE determination problem and prosumers' P2P2G trading problem.
Then the prosumers' P2P2G trading problem is further decomposed into the individual level.
At last, the P2P2G trading negotiation and DOE negotiation process are given, and the corresponding prices are analysed.
\vspace{-1.0em}
\subsection{Model reformulation and decomposition}
Having clarified the prosumers and DSO's model, (I) can be transformed into:
\begin{flalign}
  &(\mathrm{II})\underset{\left\{ P_{i,t}^{e},x_i,\forall i\in \mathcal{N} ,t\in \mathcal{T} \right\}}{\mathrm{minimize}}\,\,J\left( \left\{ \boldsymbol{P}_{i}^{e} \right\} _{i=1}^{N} \right) +\sum_{i\in \mathcal{N}}{\theta _i\left( x_i \right)}\notag&
  \\
  &\mathrm{subject}\ \mathrm{to}& \notag
  \\
  &\mathrm{For} \ \mathrm{all}\ t\in \mathcal{T} ,i\in \mathcal{N}: (\ref{batt})-(\ref{ps}), (\ref{purchase})-(\ref{prosumer_DOE}) \label{doe_prosumers2}&
  \\
  &\mathrm{For}\ \mathrm{all}\ t\in \mathcal{T} ,j\in \mathcal{L} , i\in \mathcal{\overline{N}^{+}}: (\ref{s1})-(\ref{q}) \label{doe_inj2}&
\end{flalign}
As demonstrated in (I), the DOE calculation model considering P2P2G trading necessitates the integration of both P2P2G trading and network constraints.
In order to avoid the DSO compromising prosumers' privacy, ADMM is employed to decompose (II).

The export limit ${P}_{i,t}^{e}$ is regarded as a decision variable for the $i$-th prosumer, which means prosumers are required to determine their intended export limits based on their real-time trading preferences. 
Similar to P2P trading, we formulate a reciprocity constraint requiring prosumers' intended export limits and that calculated by the DSO to be equal when negotiation terminates:
\begin{equation}
 \Psi_{i,t}: P_{i,t}^{e} = P_{i,t}^{dso,e} \label{DOE_consensus}
\end{equation}
where $P_{i,t}^{dso,e}$ denotes the DSO-side export limit.
Adding constraints (\ref{DOE_consensus}) into (II) and relaxing it using ADMM, we can decompose (II) into the following two subproblems :
\begin{small}
\begin{flalign*}
  &(\mathbf{DSO})\mathrm{minimize} \quad J\left( \left\{ \boldsymbol{P}_{i}^{dso,e} \right\} _{i=1}^{N} \right) &
  \\
  &+\sum_{t\in \mathcal{T}}\{\sum_{i\in \mathcal{N}}{\Psi _{i,t}(P_{i,t}^{dso,e}-P_{i,t}^{e})}
  +\frac{\rho}{2}\sum_{i\in \mathcal{N}}{\left( P_{i,t}^{dso,e}-P_{i,t}^{e} \right) ^2}\}\Delta T&
  \\
  &\mathrm{variables:}\{P_{i,t}^{dso,e}, f_{j,t}^{p/q}, v_{i,t}, l_{j,t}, p/q_{0,t},
  \\
  &\quad \quad \quad \quad \quad \forall t\in \mathcal{T} ,j\in \mathcal{L}, i\in \mathcal{\overline{N}^{+}}\}
  \\
  &\mathrm{subject}\ \mathrm{to}&
  \\
  &\mathrm{For}\ \mathrm{all}\ t\in \mathcal{T} ,j\in \mathcal{L}, i\in \mathcal{\overline{N}^{+}}: (\ref{s1})-(\ref{q})&
 \end{flalign*}
\end{small}

\begin{small}
\begin{flalign}
  &\mathbf{(P2P2G)}\mathrm{minimize} \sum_{t\in \mathcal{T}}{\{}\sum_{i\in \mathcal{N}}{\left( p_{i,t}^{+}\lambda _{t}^{+}-p_{i,t}^{-}\lambda _{t}^{-} \right)} \notag&
  \\
  &\quad +\sum_{i\in \mathcal{N}}{\Psi _{i,t}(P_{i,t}^{dso,e}-P_{i,t}^{e})}+\frac{\rho}{2}\sum_{i\in \mathcal{N}}{\left( P_{i,t}^{dso,e}-P_{i,t}^{e} \right) ^2}\}\Delta T\notag&
  \\
  &\mathrm{variables}:\{p_{i,t}^{+,-,inj},p_{i,t}^{batt},p_{i,t}^{P2P},E_{i,t}^{s},e_{ij,t},\notag &
  \\
  &\quad \quad \quad \quad \quad \forall i\in \mathcal{N}, j\in \mathcal{N} _i,t\in \mathcal{T}\} \notag
  \\
  &\mathrm{subject}\ \mathrm{to}& \notag
  \\
  &\mathrm{For} \ \mathrm{all}\ i\in \mathcal{N} , t\in \mathcal{T} , j\in \mathcal{N} _i:\lambda _{ij,t}:e_{ij,t}=-e_{ji,t} \label{recipro1}&
  \\
  &\quad \, \, \, \quad \quad \quad \quad \quad (\ref{batt}) -(\ref{ps}),(\ref{purchase})-(\ref{sell}),(\ref{inj})-(\ref{prosumer_DOE})\notag &
\end{flalign}
\end{small}
where $\rho$ is an empirical parameter.
In $\mathbf{(P2P2G)}$, when the optimal solution is obtained, there is $\sum_{i \in \mathcal{N}}\sum_{j\in \mathcal{N} _i}{\lambda_{ij,t}e_{ij,t}} = 0$, so the term for trading revenue in the individual objective function is eliminated in $\sum_{i\in \mathcal{N}}^{}{\theta \left( x_i \right)}$.
\vspace{-1.0em}
\subsection{COCA-based distributed algorithm for P2P2G trading}
\vspace{-0.5em}
Note that different prosumers' decision variables are coupled by the reciprocity constraint (\ref{recipro1}), taking advantage of the distributed algorithm can decompose $\mathbf{(P2P2G)}$ into the individual level.
In this paper, our algorithm is based on the consensus ADMM, advanced by a censored communication segment to reduce communication costs.

First, we introduce consensus variables into constraint (\ref{recipro1}) as:
\begin{equation}
  \lambda _{ij,t}: e_{ij,t}=-e_{ji,t}=\hat{e}_{ij,t}, \forall i\in \mathcal{N} ,\forall j\in \mathcal{N} _i \label{consensus}
\end{equation}
the consesus variable $\hat{e}_{ij,t}$ can be regarded as local trading amount estimation of prosumer $i$.

Then we use Lagrange multipliers of constraint (\ref{consensus}) to formulate the augmented Lagrangian, after which the subproblem for each prosumer can be written as:
\begin{flalign*}
 &\mathbf{(Prosumer)}\mathrm{minimize}\ \sum_{t\in \mathcal{T}}{\Delta T\{p_{i,t}^{+}\lambda _{t}^{+}-p_{i,t}^{-}\lambda _{t}^{-}}&
 \\
 &\quad \quad \quad \quad \quad +\sum_{j\in \mathcal{N} _i}{\lambda _{ij,t}\left( \hat{e}_{ij,t}-e_{ij,t} \right)}+\frac{\rho}{2}\sum_{j\in \mathcal{N} _i}{\left( \hat{e}_{ij,t}-e_{ij,t} \right) ^2}&
 \\
 &\quad \quad \quad \quad \quad +\Psi _{i,t}(p_{i,t}^{'DOE}-p_{i,t}^{DOE})+\frac{\rho}{2}(p_{i,t}^{'DOE}-p_{i,t}^{DOE})^2\}&
 \\
 &\mathrm{variables:}\{P_{i,t}^{e}, p_{i,t}^{+,-,inj},p_{i,t}^{batt},p_{i,t}^{P2P},E_{i,t}^{s},e_{ij,t},&
 \\
 &\quad \quad \quad \quad \quad \forall j\in \mathcal{N} _i,t\in \mathcal{T}\}
 \\
 &\mathrm{subject}\ \mathrm{to}&
 \\
 &\mathrm{For}\ \mathrm{all}\ t\in \mathcal{T} ,j\in \mathcal{N} _i: (\ref{batt}) -(\ref{ps}),(\ref{purchase})-(\ref{sell}),(\ref{inj})-(\ref{prosumer_DOE})&
\end{flalign*}
In the $k$-th iteration, prosumer $i$ solves local optimization problem $\mathbf{(Prosumer)}$ based on the local estimations of trading amount $\hat{e}_{ij,t}^{k-1}$ and price $\lambda_{ij,t}^{k-1}$, then it obtains the updated P2P trading amount $e_{ij,t}^{k}$ with its partner $j$.
At the same time, prosumer $j$ calculates and sends its corresponding decision $e_{ji,t}^{k}$ to prosumer $i$. 
Thereupon, prosumer $i$ can update the local estimations of trading amount and price as follows, which coincides with the updating rules of ADMM:
\begin{small}
\begin{align}
  &\hat{e}_{ij,t}^{k}=\frac{1}{2}\left( e_{ij,t}^{k,*}-e_{ji,t}^{k,*} \right) \label{pv_update}\\
  &\lambda _{ij,t}^{k}=\lambda _{ij,t}^{k-1}+\rho \left( \hat{e}_{ij,t}^{k,*}-e_{ij,t}^{k,*} \right) \label{dv_update}
\end{align}
\end{small}
where $\rho$ is an empirical parameter. The update continues until the algorithm converges, as the total residuals fall below the criteria:
\begin{small}
\begin{equation}
\begin{aligned}
  &\sum_{i\in \mathcal{N}}{\sum_{j\in \mathcal{N} _i}{\sum_{t\in \mathcal{T}}{\left( e_{ij,t}^{k}+e_{ji,t}^{k} \right) ^2\,\,}}}\le \,\,\chi _{e}^{s} \\
  &\sum_{i\in \mathcal{N}}{\sum_{j\in \mathcal{N} _i}{\sum_{t\in \mathcal{T}}{\left( e_{ij,t}^{k}-e_{ij,t}^{k-1} \right) ^2\,\,}}}\le \,\,\chi _{e}^{t} \label{trades_convergence_primal}
\end{aligned}
\end{equation}
\end{small}
where $\chi _{e}^{s/t}$ is the presetted threshold.

However, the proposed consensus ADMM method may lead to a heavy communication burden, especially when a lot of iterations are required. 
By adding a censored strategy into the communication process, the COCA algorithm can effectively reduce communication costs without compromising the convergent rate.
Applying the strategy in P2P2G trading, a prosumer is not allowed to send its updated trading amounts to partners if they are not sufficiently different from the previous one.
The threshold for communication in $k$-th iteration can be described as:
\begin{align}
  &\xi _{i}^{k}=\hat{e}_{i}^{k-1}-e_{i}^{k} \label{diff}\\
  &H_i\left( k,\xi _{i}^{k} \right) =\left\| \hat{e}_{i}^{k-1}-e_{i}^{k} \right\| -\alpha m ^{k}\,\,\geq \,\,0  \label{threshold}
\end{align}
where $\xi _{i}^{k}$ measures the change of the locally derived trading amount of prosumer $i$, and $H_i$ compares such change with adaptive threshold $m^{k}$.
If (\ref{threshold}) is satisfied, prosumer $i$ can communicate with its partners and vice versa. 
By properly choosing threshold $m$, the convergence of the algorithm can be guaranteed \cite{COCA}.

In summary, in the proposed P2P2G trading model, prosumers only need to communicate their updated P2P trading amount with each other, and P2P prices are estimated locally.
Therefore, the mechanism is totally decentralized. 
Combined with a communication censored strategy, prosumers' communication costs can also be saved compared to most distributed optimization-based methods.
The total cost of prosumers obtained by this method also turns out to be the same as that of the centralized problem, which is proved in \cite{COCA}.
\subsection{ADMM-based DOE negotiation}
In this subsection we depict the DOE negotiation process.
Take a panoramic view of the mechanism, in $k$-th iteration, prosumer $i$ solves $\mathbf{(Prosumer)}$ to submit $P_{i,t}^{e}$ to the DSO and P2P trading amounts $e_{ij,t}$ to its potential P2P trading partner $j$.
According to others' trading response, prosumer $i$ updates $\hat{e}_{ij,t}$ and $\lambda_{ij,t}$. 
Simultaneously, the DSO solves $\mathbf{(DSO)}$ to calculate $P_{i,t}^{dso,e}$ and update $\Psi_{i,t}$ by:
\begin{equation}
  \Psi _{i,t}^{k}=\Psi _{i,t}^{k-1}+\rho ({P_{i,t}^{dso,e^{k,*}}}-P_{i,t}^{e^{k,*}}) \label{DOE_update}
\end{equation}

\textit{Remark 2:} It is difficult to derive network functions (\ref{s1})-(\ref{q}) in closed form, so this nonconvex model $\mathbf{(DSO)}$ cannot be easily solved.
Many methods such as second-order conic programming (SOCP) relaxation, semi-definite programming (SDP) relaxation can be used to solve it, and they are proven to be exact under certain conditions.
In this paper, we use the SOCP relaxation of the DistFlow model to solve the above optimization problem, and the result is shown to be exact.
The network variables like voltage magnitudes in the case study are derived from solving the SOCP relaxed model.
But in price analysis, we do not base on the SOCP model to illustrate because the formulation of DOE price is not that interpretable compared with that obtained from the model using implicit function constraints.

After the DSO broadcasting $P_{i,t}^{dso,e}$ and $\Psi_{i,t}$ to prosumers, the next iteration begins and repeats until convergence.
It is crucial to mention that two negotiations influence each other, as whenever the DSO issues updated DOEs and price signals, by refreshing DOE constraints, prosumers will fine-tune their P2P2G trading behaviors to satisfy DOE limits.

The overall pseudocode of the proposed market mechanism for P2P2G trading is presented in Algorithm \ref{algo:algo1}.
Algorithm \ref{algo:algo1} converges when (\ref{trades_convergence_primal}) is satisfied along with following inequalities:
\begin{small}
\begin{equation}
  \begin{aligned}
  &\sum_{i\in \mathcal{N}}{\sum_{t\in \mathcal{T}}{\left( P_{i,t}^{dso,e^k}+P_{i,t}^{e^k} \right) ^2\,\,}}\le \,\,\chi _{d}^{s} \\
  &\sum_{i\in \mathcal{N}}{\sum_{t\in \mathcal{T}}{\left( P_{i,t}^{dso,e^k}-P_{i,t}^{dso,e^k} \right) ^2\,\,}}\le \,\,\chi _{d}^{t} \label{DOE_convergence_primal}
  \end{aligned}
\end{equation}
\end{small}
where $\chi _{d}^{s/t}$ is presetted threshold.

\begin{algorithm}
	\caption{DOE Embedded P2P2G Market Clearing}
	\label{algo:algo1}
  \IncMargin{1em}
	\KwIn{For all prosumers $i\in\mathcal{N}$, $j\in\mathcal{N}_i$, $t\in\mathcal{T}$: 
  $\hat{e}_{ij,t}^{0}$, $\lambda_{ij,t}^{0}$, $P_{i,t}^{dso,e^{0}}$, $\Psi_{i,t}^{0}$,
  and $m^{0}$, $\alpha$, $\rho$.}
	\KwOut{$e_{ij,t}$, $\lambda_{ij,t}$, $P_{i,t}^{dso,e}$, $P_{i,t}^{e}$, $\Psi_{i,t}$,}
	
  Iterate untill convergence ($k=0,1,\dots$): (\ref{trades_convergence_primal}), (\ref{DOE_convergence_primal})\\

  \textbf{Prosumers:}

  \For{all prosumers $i\in\mathcal{N}$ (in parallel)}{
    Solve $\mathbf{(Prosumer)}$:	\textit{local determination.}\\

    Communication-censored segment: (\ref{diff}), (\ref{threshold}).\\
    \eIf{(\ref{threshold}) satisfies:}{
      Send $e_{ij,t}^{k}$ to $j\in\mathcal{N}_i$ respectively.
      }{Do not send $e_{ij,t}^{k}$.}

    \eIf{Receive $e_{ji,t}^{k}$ from $j\in\mathcal{N}_i$}{
      Update local trading amounts and price: (\ref{pv_update}), (\ref{dv_update}).
      }{Update (\ref{pv_update}), (\ref{dv_update}) using $e_{ji,t}^{k}=e_{ji,t}^{k-1}$.}
    
    Send intended export limit $P_{i,t}^{e^{k}}$ to DSO.\\

    Receive $P_{i,t}^{dso,e^{k}}$, $\Psi_{i,t}^{k}$ from DSO.\\

    Update communication-censored threshold.
}

  \textbf{DSO:}

    Receive prosumer-side export limit $P_{i,t}^{e^{k}}$ from $i\in\mathcal{N}_i$.\\

    Solve $\mathbf{(DSO)}$:	\textit{security constrained OPF}.\\

    Update DOE price: (\ref{DOE_update}).\\

    Broadcast $P_{i,t}^{dso,e^{k}}$, $\Psi_{i,t}^{k}$ to prosumers.

\end{algorithm}
\vspace{-1.0em}
\subsection{Components of DOE and P2P Price}
\vspace{-0.5em}
Assuming that we have already found the optimal solution, from Envelope Theorem we can get the DOE price as:
\begin{small}
\begin{align}
  \frac{\partial \mathcal{L} _{DSO}}{\partial P_{i,t}^{e}}&=-\Psi _{i,t} \label{envelop_th}
\end{align}
\end{small}
where $\mathcal{L}_{DSO}$ is the Lagrangian of the DSO subproblem $\left( \mathbf{DSO} \right) $, and $\Psi$ is the lagrange multiplier of the DOE reciprocity constraint (\ref{DOE_consensus}).
Next, we illustrate the components of such DOE price.

From KKT condition, we can obtain $\frac{\partial \mathcal{L} _{DSO}}{\partial P_{i,t}^{dso,e}}=0$.
Combined with (\ref{envelop_th}), we can show the DOE price as:
\begin{scriptsize}
\begin{align*}
	&\frac{\partial \mathcal{L} _{DSO}}{\partial P_{i,t}^{e}}=-\Psi _{i,t}=\rho \left( P_{i,t}^{dso,e}-P_{i,t}^{e} \right)\\
	&+\Sigma _s\frac{s}{S}\{\Sigma _j\eta _{j,t,s}(2f_{j,t}^{p}\frac{\partial f_{j,t,s}^{p}}{\partial P_{i,t}^{dso,e}}+2f_{j,t}^{q}\frac{\partial f_{j,t,s}^{q}}{\partial P_{i,t}^{dso,e}})\\
	&+\Sigma _j\Delta _{j,t,s}[2(f_{j,t,s}^{p}\frac{\partial f_{j,t,s}^{p}}{\partial P_{i,t}^{dso,e}}+f_{j,t,s}^{q}\frac{\partial f_{j,t,s}^{q}}{\partial P_{i,t}^{dso,e}}-(R_j+X_j)\frac{\partial l_{j,t,s}}{\partial P_{i,t}^{dso,e}})]\\
	&+\Sigma _i(\tau _{i,t,s}^{+}-\tau _{i,t,s}^{-})\frac{\partial v_{i,t,s}}{\partial P_{i,t}^{dso,e}}\\
	&+\Omega _{p,t,s}\\
	&+(\Omega _{p,t,s}+\frac{\pi _t}{S})\Sigma _jR_j\frac{\partial l_{j,t,s}}{\partial P_{i,t}^{dso,e}}+\Omega _{q,t}\Sigma _jX_j\frac{\partial l_{j,t,s}}{\partial P_{i,t}^{dso,e}}\}\\
\end{align*}
\end{scriptsize}
After Algorithm \ref{algo:algo1} converges, the first term on the right-hand side in the above equation is equal to zero. 
The second and third terms are related to the lines' power limits.
The fourth term is related to voltage magnitudes limits. The fifth term is related to power at the root. The last two terms are related to network power loss.

Expressing $\Psi_{i,t}$ in this way, it can be transparently interpreted as the sum of the congestion component, voltage component, energy component and loss component.
Passing such a DOE price signal to prosumers is reasonable, as it indicates the potential marginal cost for the DSO to maintain the network integrity of the distribution network when prosumers ask for a larger DOE at its node.

Furthermore, we can calculate the Lagrangian of the prosumer subproblem  $\left( \mathbf{Prosumer} \right) $, and use KKT condition to deduce:
\begin{equation}
  \begin{aligned}
    \lambda _{ij,t}&=-\mu _{i,t}=-\phi _{i,t}-\epsilon _{i,t}=-\phi _{i,t}-\gamma _{i,t}\\
    &=-\phi _{i,t}+\Psi_{i,t}\\
  \end{aligned}
  \end{equation}
From the fourth equation, we can see that the prosumer's P2P trading price is internally influenced by the DOE price.
Therewith, prosumers' trading behaviors will be guided to be grid-friendly.

\vspace{-1.0em}
\section{Case Study}\label{sec:case}
\subsection{Simulation Setting}
All numerical studies were performed in Python and used the Gurobi solver.
Actual power flow after P2P2G trading was simulated using PyPSA \cite{PyPSA} in Python.
A 15-bus test system in Fig. \ref{fig:15line} was used to analyze the proposed P2P2G trading mechanism.
The network parameters were taken from reference \cite{DLMP_Dominant_GUP}. 
We assumed only one prosumer locates at Node 2,7 and 12, and they participate in P2P2G trading and need to obey DOE limits. 
All remaining nodes were viewed as consumer nodes with fixed active and reactive loads. 
In terms of prices, we set ToU twice the value of FiT. 
With respect to algorithm parameters, we set the initial P2P trading price to the average of FiT and ToU, and initial trading amounts are set zero. 
Initial DOE is set larger than the maximum RES output.
The convergence thresholds in stopping criteria of the algorithm were set $1.5\times 10^{-5}$.

\subsection{15-bus system test}
Without considering DOE, the P2P2G problem solution takes 93 iterations to converge, and the result shows voltages and line powers are out of limits.
After incorporating DOE, the proposed mechanism solution is obtained after 187 iterations, and the network integrity problems are eliminated.
Next, we compare these two results and reveal the superiority of the proposed mechanism.

1) \textit{Impact of DOEs on the network integrity:}
In this part, we first compared the network integrity before and after incorporating DOEs into P2P2G trading.

\begin{figure}[htb]
  \centering
  \includegraphics[width=0.95\columnwidth]{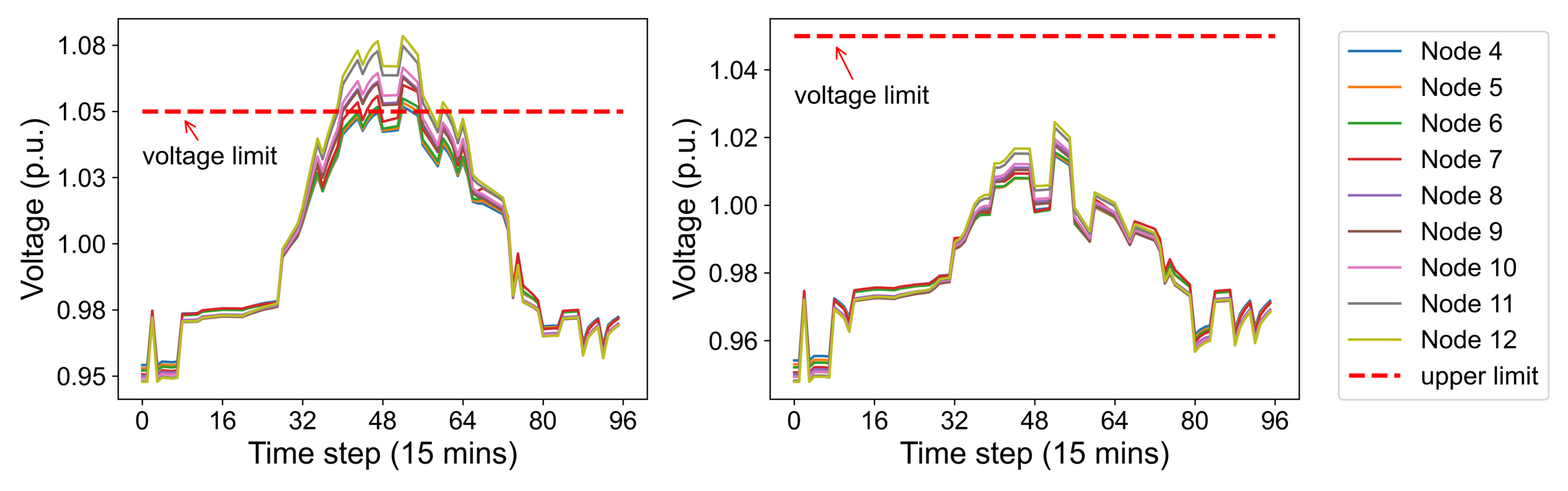}
  \caption{Comparison of voltage profile w/o (Left) and w/ DOE (Right).}
  \label{fig:15v}
  \end{figure}

Fig. \ref{fig:15v} shows the voltage profile after P2P2G trading without consideration of network constraints.
Due to unlimited trading, there are problems of voltage limit violations, especially for those nodes far from the root.
After introducing DOEs into P2P2G trading, the voltages are within a safe range.

Also, before using DOEs, the network power loss is high, because frequent P2P trades inject more unexpected power into the grid.
Regulated by the DSO through DOEs, the loss of energy during the whole day in the distribution network reduces from 1.175MWh to 0.245MWh.

\begin{figure}[tb]
  \centering
  \includegraphics[width=0.95\columnwidth]{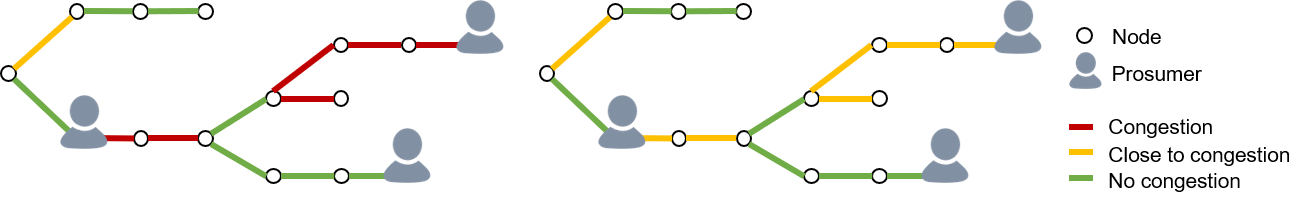}
  \caption{Comparison of congestion status w/o (Left) and w/ DOE (Right).}
  \label{fig:15line}
  \end{figure}

As for line congestion, the line statuses before and after using DOEs are illustrated in Fig. \ref{fig:15line}.
Not taking DOEs into consideration, we observed that several lines suffer from congestion, which is harmful to the grid and will cause further maintenance.
However, after using DOEs, the congestion is eliminated.

\begin{figure}[tb]
  \centering
  \includegraphics[width=0.95\columnwidth]{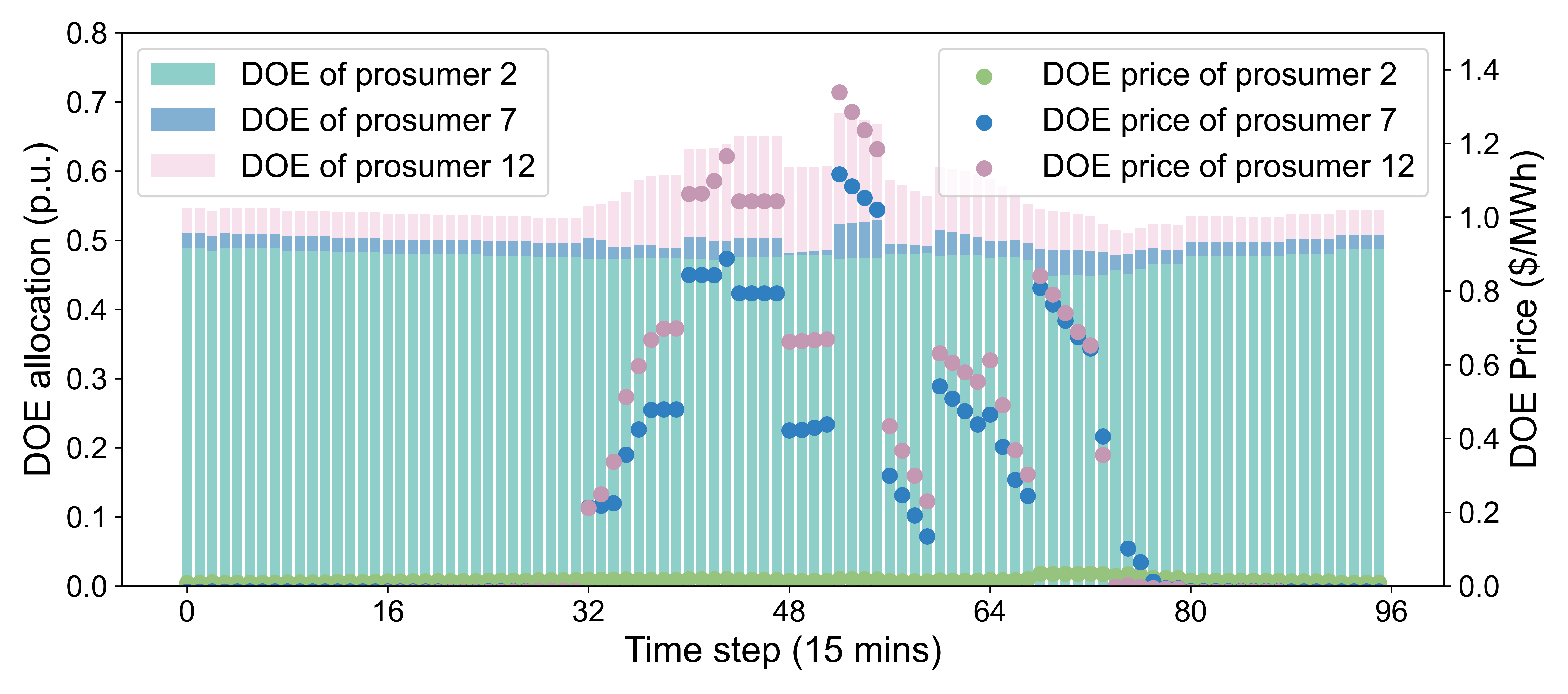}
  \caption{DOE allocation and DOE prices for prosumers.}
  \label{fig:15DOE}
  \end{figure}

The negotiated DOE is shown as bars in Fig. \ref{fig:15DOE}. The result follows the intuition that the furthest prosumer suffers the most severe network problem.
So, prosumer 7 and 12 who are located at the terminal of this radial network are allowed the thinnest DOE, while prosumer 2 close to the root enjoys the most.

All results above imply that when the DOE is incorporated into P2P2G trading, trading behaviors are regulated, thus network integrity problems are eliminated.

\begin{figure}[htb]
  \centering
  \includegraphics[width=0.95\columnwidth]{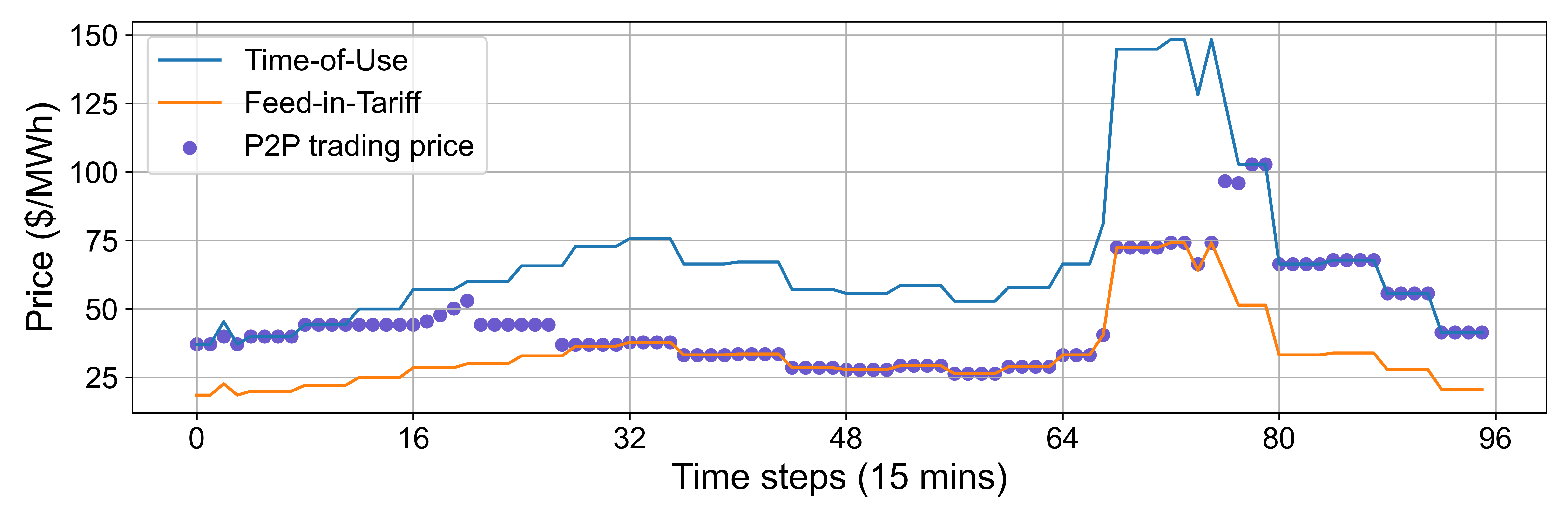}
  \caption{P2P price of prosumers.}
  \label{fig:15p2pprice}
  \end{figure}

2) \textit{Economical Performance:}
As shown in Fig. \ref{fig:15p2pprice}, the P2P trading price lies between FiT and ToU, which can motivate prosumers to trade with peers.
The result shows that the surplus of prosumers increases from 101.73\$ in P2G trading only to 126.38\$ after P2P trading.
Thus, prosumers can collectively reduce their total purchase cost from the distribution network.

We also plotted the DOE price of three prosumers in Fig. \ref{fig:15DOE}. Consistent with the results of DOE allocation, the prosumer at the terminal node is asked for the most expensive DOE charge, 
which means DOE requirement of this node will lead to more costs for the DSO to keep distribution network integrity than other nodes.

\begin{figure}[htb]
  \centering
  \includegraphics[width=0.95\columnwidth]{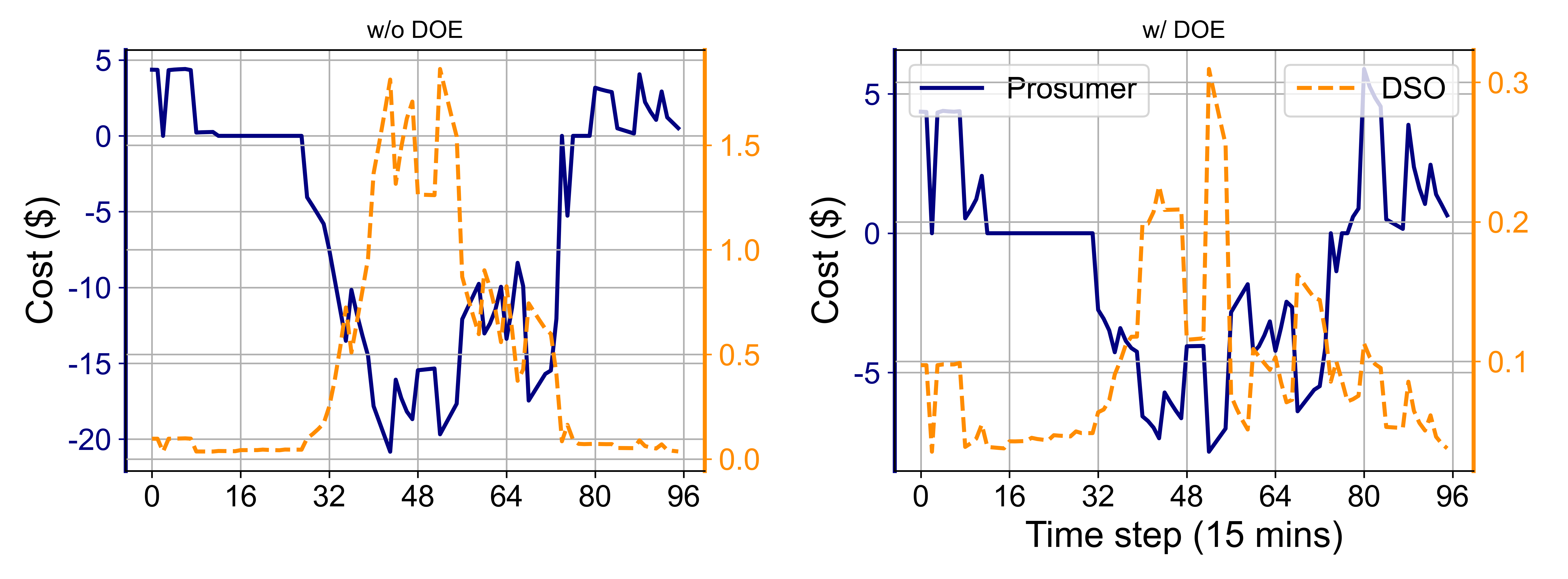}
  \caption{Comparison of cost profile w/o (Left) and w/ DOE (Right).}
  \label{fig:15cost_t}
  \end{figure}

Fig. \ref{fig:15cost_t} shows the cost profile of prosumers and the DSO respectively before and after using DOEs.
It is obvious that without DOEs, the DSO's operation cost is high, especially in the mid of the day.
The reduction of power loss lightens that burden for the DSO.

3) \textit{Computational Performance:}
To demonstrate the feasibility of the proposed P2P2G trading mechanism, we illustrated the computational performance of the 15-bus case study.
\begin{figure}[htb]
  \centering
  \includegraphics[width=0.95\columnwidth]{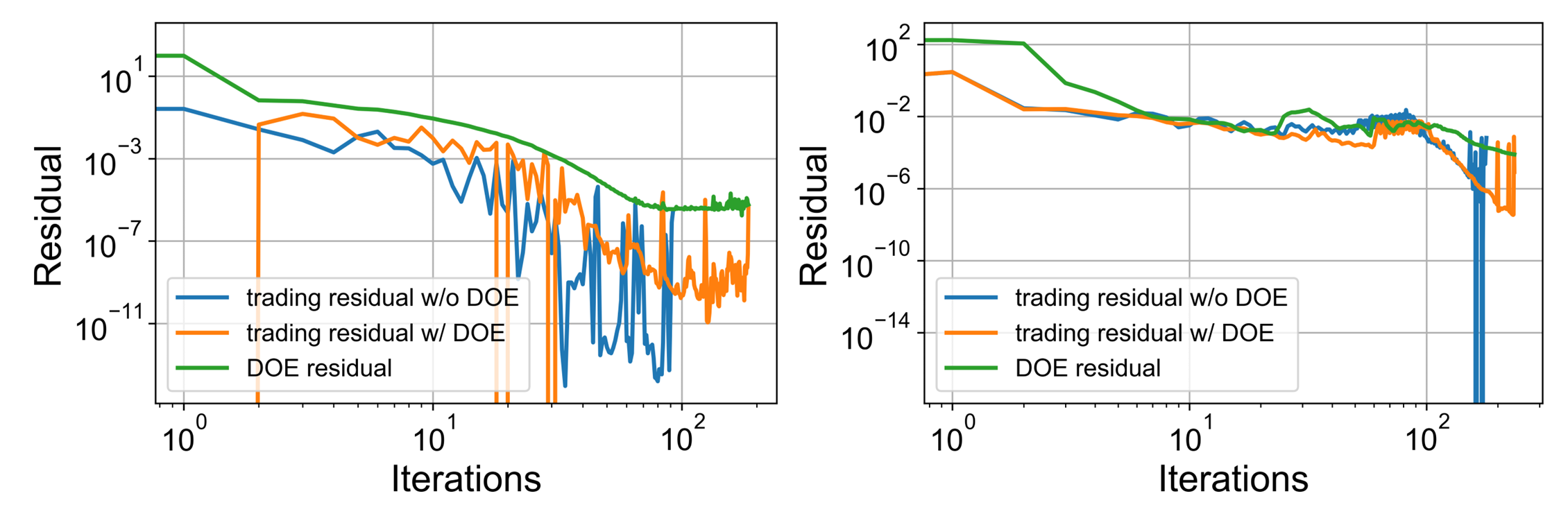}
  \caption{Convergence in 15-bus test (Left) and 141-bus test (Right)}
  \label{fig:conver}
  \end{figure}

Primal residuals versus iteration are plotted in Fig. \ref{fig:conver}.
As in each iteration, there are actually two ADMM-based negotiations carrying on, if one negotiation terminates much later than the other, the other one may conduct many redundant communications and lead to an unnecessary transmission burden.
From Fig. \ref{fig:conver}, the convergence rate of DOE negotiation is much slower than P2P trading.
Even though prosumers need to adjust trading behaviors to follow DOE negotiation, in abundant iterations the change of each prosumer's trading amounts is slight.
Taking advantage of COCA, in many iterations, prosumers need not send trading amounts to others.
In the 15-bus system test, instead of using COCA, traditional ADMM takes 187 iterations to converge and overall P2P communication times are 1122.
Replacing ADMM with COCA, the process also takes 187 times to converge, but the total P2P communication times are merely 860.
Therefore, the communication cost for each prosumer is significantly reduced.
\vspace{-1.0em}
\subsection{141-bus system test}
To show that the proposed mechanism scales well with larger networks and more prosumers, we conducted the simulation on a 141-bus system.
In this case, we ran for 24 hours and arranged 28 prosumers randomly at different nodes. RES outputs and storage system parameters are also set differently from each other.
Before using DOEs, the 141-system mainly suffers from voltage over limits problems.
For quicker convergence, we set convergence thresholds $1\times 10^{-3}$.
Primal residuals versus iteration are plotted in Fig. \ref{fig:conver}.

The simulation result is shown in Table \ref{tab:141res}.
\begin{table}[h]
\vspace{-1.0em}
  \small
  \caption{141-bus test system results} \label{tab:141res}
  \resizebox{\columnwidth}{!}{%
  \begin{tabular}{|c|cc|c|c|}
  \hline
  \multirow{2}{*}{Approach} & \multicolumn{2}{c|}{\#Overlimits}                                                                                                                   & \multirow{2}{*}{\begin{tabular}[c]{@{}c@{}}Total loss \\ (MWh)\end{tabular}} & \multirow{2}{*}{\begin{tabular}[c]{@{}c@{}}\#Communication \\ with peers\end{tabular}} \\ \cline{2-3}
                            & \multicolumn{1}{c|}{\begin{tabular}[c]{@{}c@{}}Voltage \\ (\#nodes)\end{tabular}} & \begin{tabular}[c]{@{}c@{}}Line power \\ (\#lines)\end{tabular} &                                                                              &                                                                                        \\ \hline
  M1: P2P2G trading without DOE & \multicolumn{1}{c|}{45}                                                           & 11                                                              & 9633                                                                         & 136836                                                                                 \\ \hline
  M2: DOE + P2P2G with ADMM   & \multicolumn{1}{c|}{0}                                                            & 0                                                               & 6648                                                                         & 179172                                                                                 \\ \hline
  M3: DOE + P2PG with COCA    & \multicolumn{1}{c|}{0}                                                            & 0                                                               & 6648                                                                         & 20731                                                                                 \\ \hline
  \end{tabular}%
  }
\end{table}

\begin{figure}[htb]
  \centering
  \includegraphics[width=0.95\columnwidth]{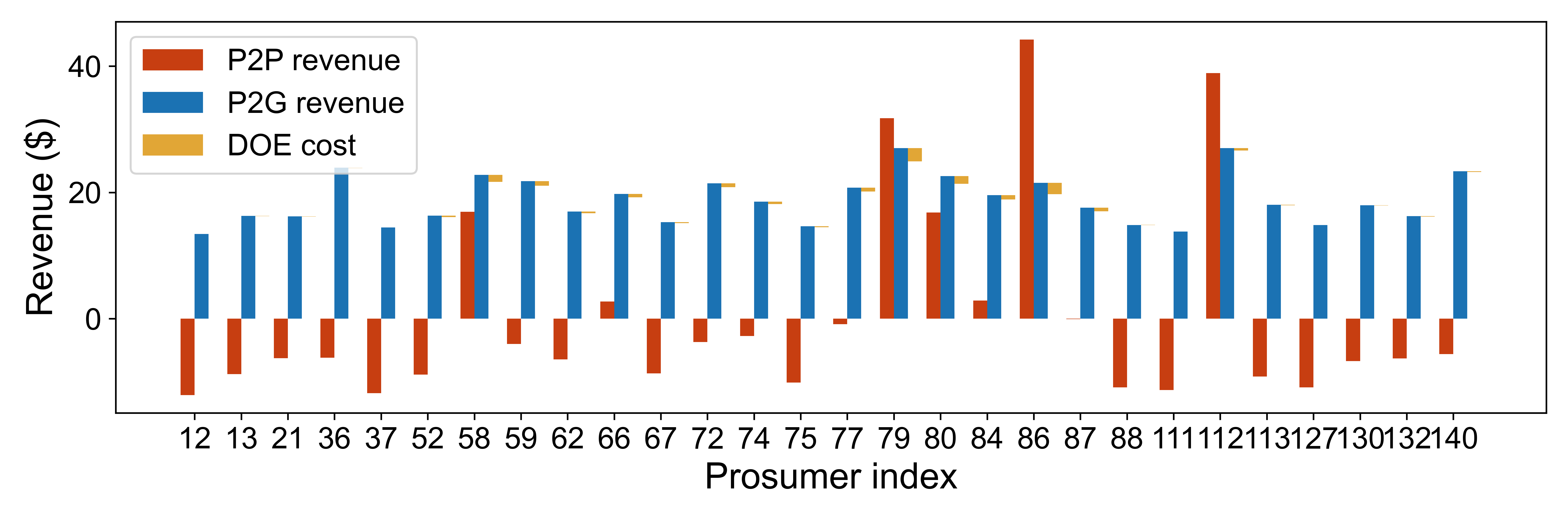}
  \caption{Prosumers' revenues of 141-system.}
  \label{fig:141cost}
\end{figure}

From the comparison between M1 and M2/M3, we can see that the voltage and line power problems happening in the P2P2G trading are eliminated, and the total energy loss of the distribution network is reduced by one-third.
From Fig. \ref{fig:141cost}, prosumers' revenues are shown.
With a small ratio of DOE payment, the DSO can use DOE to direct prosumers to be grid-friendly.

Embedded with DOE, the M2/M3 takes more iterations to converge than M1, because there is DOE negotiation added, the communication times among peers increase from 136836 to 179172.
However, many of them are unnecessary when DOE negotiation is far from the end.
To deal with the communication burden, making use of the censored strategy, DOE embedded P2P2G trading only needs 20731 communication times among peers in M3 to carry on P2P trading negotiation, even less than that of M1.

\section{Conclusion}\label{sec:conclusion}
This work proposes a novel P2P2G mechanism considering distribution network integrity.
By introducing DOEs, network integrity under P2P trading among prosumers can be guaranteed without the DSO interfering trading process.
Solving by ADMM-based algorithm, DOEs are determined during negotiations between the DSO and prosumers, preserving prosumers' privacy well.
A COCA algorithm used in the P2P trading process reduces prosumers' communication costs, which makes the mechanism more practical.
Finally, DOE price and P2P trading price determination in this paper are shown to have an economical interpretation.

This work assumes prosumers have non-flexible loads and the mechanism works in the day-ahead market.
Future research should focus on considering flexible loads, and try to apply the mechanism under a real-time condition.
Some communication-efficient distributed algorithms like asynchronous ADMM can also be tested to avoid more communication problems.

\bibliographystyle{IEEEtran}
\bibliography{main}

\end{document}